\documentclass{IEEEtran4PSCC}

\usepackage{cite}

\ifCLASSINFOpdf
   \usepackage[pdftex]{graphicx}
\else
   \usepackage[dvips]{graphicx}
\fi
\usepackage{xcolor}

\usepackage[cmex10]{amsmath}
\ifCLASSOPTIONcompsoc
  \usepackage[caption=false,font=normalsize,labelfont=sf,textfont=sf]{subfig}
\else
  \usepackage[caption=false,font=footnotesize]{subfig}
\fi
 \usepackage{url}
\makeatletter
\let\old@ps@headings\ps@headings
\let\old@ps@IEEEtitlepagestyle\ps@IEEEtitlepagestyle
\def\psccfooter#1{%
    \def\ps@headings{%
        \old@ps@headings%
        \def\@oddfoot{\strut\hfill#1\hfill\strut}%
        \def\@evenfoot{\strut\hfill#1\hfill\strut}%
    }%
    \def\ps@IEEEtitlepagestyle{%
        \old@ps@IEEEtitlepagestyle%
        \def\@oddfoot{\strut\hfill#1\hfill\strut}%
        \def\@evenfoot{\strut\hfill#1\hfill\strut}%
    }%
    \ps@headings%
}
\makeatother

\psccfooter{%
        \parbox{\textwidth}{\hrulefill \\ \small{Accepted to the 21st Power Systems Computation Conference (PSCC 2020)} }%
}

\allowdisplaybreaks

\begin{document}
%
\title{Data Driven Transfer Functions and Transmission Network Parameters for GIC Modelling}
 \author{\IEEEauthorblockN{M. J. Heyns\IEEEauthorrefmark{1}\IEEEauthorrefmark{2},
 C. T. Gaunt\IEEEauthorrefmark{1},
 S. I. Lotz\IEEEauthorrefmark{2} and
 P. J. Cilliers\IEEEauthorrefmark{1}\IEEEauthorrefmark{2}}
 \IEEEauthorblockA{\IEEEauthorrefmark{1} Department of Electrical Engineering\\
 University of Cape Town (UCT),
 Cape Town, South Africa}
 \IEEEauthorblockA{\IEEEauthorrefmark{2} South African National Space Agency (SANSA)\\
 Hermanus, South Africa
 \\mheyns@sansa.org.za}
 }
\maketitle
\begin{abstract}
Typical geomagnetically induced current (GIC) modelling assumes the induced quasi-DC current at a node in the transmission network is linearly related to the local geoelectric field by a pair of network parameters. Given a limited time-series of measured geomagnetic and GIC data, an empirical method is presented that results in a statistically significant generalised ensemble of parameter estimates with the error in the estimates identified. The method is showcased for different transmission networks and geomagnetic storms and, where prior modelling exists, shows improved GIC estimation. Furthermore, modelled networks can be locally characterised and probed without any further network knowledge. Insights include network parameter variation, effective network directionality and response. Merging the network parameters and geoelectric field estimation, a transfer function is derived which offers an alternative approach to assessing transformer exposure to GICs.
\end{abstract}
\begin{IEEEkeywords}
Ensemble estimation, geomagnetically induced currents (GICs), transmission network parameters
\end{IEEEkeywords}
\thanksto{\noindent This work was supported in part by a grant from the Open Philanthropy Project. The authors would like to further acknowledge Eskom and the EPRI Sunburst project for measured GIC data in South Africa and similarly Powerlink Queensland in Australia and Tennessee Valley Authority in the USA, for which CTG is a strategic partner.}
\section{Introduction}
The effects of geomagnetically induced currents (GICs) in communications and power systems were well known for many years before the first significant papers on calculating the GICs appeared \cite{Albertson1970}. Initially, the driving near-Earth current system was modelled as a line or sheet, giving by first principles different answers for the power line currents. Another approach gave the transformer neutral currents directly - by calculating the DC-equivalent of the voltages induced in the whole network by a uniform plane-wave \cite{Lehtinen1985a}. For a given node, the traditional nodal modelling formulation is
\begin{align}
GIC(t)= aE_x(t) + bE_y(t),\label{eqn:trad}
\end{align}
where $a$ and $b$ are derived constants based on network topology and resistances assuming 1 V/km geoelectric (E-field) components $E_x$ and $E_y$, where $x$ and $y$ indicates the North and East directions respectively. These E-field components are typically not measured, but rather derived from the measured geomagnetic field (B-field). The E-field and B-field can be related in the frequency domain through the surface impedance. Surface impedance models have varying degrees of complexity and of the several methods developed, a multi-layered ground conductivity is widely used due to its generality, simplicity and apparent accuracy \cite{Sun2019}. Regardless of E-field derivation method, assumptions of DC-equivalence and constant network parameters remain in GIC modelling approaches. A recent paper \cite{Weigel2019} examined the DC-equivalence assumption by comparing measured E-field and GIC data at a node, showing that empirical $a$ and $b$ parameters as defined by \eqref{eqn:trad} in the frequency domain are frequency dependent. The source of the frequency dependence is difficult to pin down since the measured E-field at a single node is not necessarily the same as the network effective E-field. This current work follows the nodal formulation of \eqref{eqn:trad}, but differs significantly in that the network parameters are not assumed to be constant. Relaxing this assumption allows for a simple approach of a fast, effective transform from B-field to GIC with accurate effective network parameter estimation, while acknowledging possible unmodelled frequency dependence and other uncertainty, directly applicable to planning and operations.

Building from the nodal formalism of GIC modelling, the ensemble methodology of network parameter estimation is presented in Section \ref{sec:ens}. Three different datasets from around the world are used to generate network parameter ensembles and test their performance in Section \ref{sec:res}, with the data used described in Section \ref{sec:dat}. Section \ref{subsec:char} looks specifically into the characteristics of GICs in the local networks derived from the network parameter ensembles. Section \ref{subsec:tf} further expands the ensemble methodology to compute transfer functions straight from B-field to GIC. Both the E-field to GIC and B-field to GIC ensemble methods are tested in Section \ref{sec:mod}. The focus throughout this paper is on operational modelling, with emphasis on estimating the uncertainty associated with traditional modelling.

\section{Ensemble Estimation}\label{sec:ens}
Traditional GIC modelling recognizes three steps in calculating transformer neutral currents: derivation of local B-field components from suitable measured or interpolated magnetic measurements; a frequency dependent transform through the surface impedance to an E-field; and network analysis. Assuming the driving disturbance B-field from near-Earth current systems is spatially uniform and vertically incident at the Earth's surface, along with laterally homogeneous ground conductivity, a conservative E-field would be produced through Faraday's law. Given a purely resistive network, the system can be modelled perfectly by \eqref{eqn:trad}. Although these assumptions can be justified as approximations, there are several challenges to traditional GIC modelling. Ground conductivity is not laterally homogeneous and interfaces, such as the coast, can have a significant effect on the magnitude and direction of the induced E-field \cite{Lucas2018}. Adding that the driving current systems do not produce a uniform B-field over an area the size of a network, the E-field is not strictly conservative and the transmission line shape is significant \cite{Sun2019}. The network analysis needed to derive the network parameters is not trivial and the entire network needs to be taken into account \cite{Overbye2013}. Individual transformers can influence each other \cite{Divett2018}, along with different voltage levels and possible non-linear inductive network response which is the topic of very recent research \cite{Chisepo,Chisepo2}. There are also variables such as the effect of rainfall on grounding resistance that add unmodelled complexity \cite{Blake2018b}. Given the multitude of higher-order effects, the current state of the art, which makes use of dense electromagnetic surveys and requires detailed network information, still does not fully model the nature of measured GICs. In comparison the traditional modelling framework does surprisingly well, especially when empirically `tuned' via network parameters or similar scaling factors, and is ideal for operational application.

With more and more measured GIC data, the robustness of the traditional modelling approach can be leveraged using data driven approaches. Using measured GIC data overcomes the fact that the entire network aggregates the driving variables. The resulting integrated effective response, which may arise from complex interactions, can be very different from modelling the actual driving variables at the single point of interest. Using such an empirical methodology also allows a natural error range to be assigned to each step in the modelling process, or in certain cases the cumulative error of multiple steps. The ensemble method described in this paper makes use of actual measurements of GICs, measured B-field and derived E-field, but can be applied similarly to measured E-field if available. Firstly, \eqref{eqn:trad} is updated to acknowledge the accumulated errors involved in traditional GIC modelling and data driven models in general \cite{Wik2008}. Now,
\begin{align}
GIC(t)+GIC(t)_{err}=&a\big(E_x(t)+{E_x}(t)_{err}\big)+\nonumber\\
&b\big(E_y(t)+{E_y}(t)_{err}\big)\\
\text{ or }\Gamma(t)\approx&\alpha{E_x}(t)+\beta{E_y}(t),\label{eqn:gic}
\end{align}
where $X_{err}$ indicates the error made in the measurement or estimation of parameter $X$ and $\Gamma(t)$ is the GIC as measured.

\begin{figure}[!ht]
\centering
\includegraphics[width=3.49in]{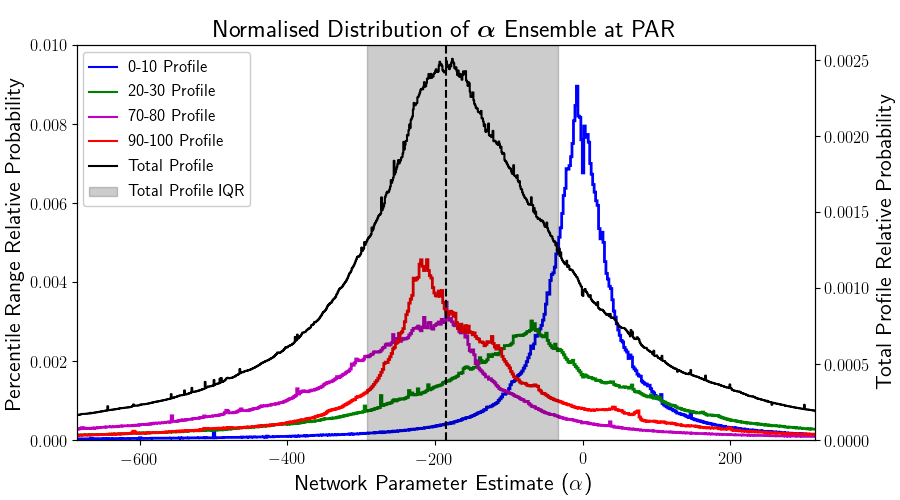}
\caption{Ensemble of $\alpha$ network parameter at PAR. Since the effective network is approximately north-south, $\beta$ is negligible in comparison. Variation and change of network parameters with GIC strength is seen in the percentile range profiles. The total ensemble parameter estimate is indicated by the dashed line, with the associated error signified by the interquartile range (IQR).}
\label{fig_ens}
\end{figure}

Simultaneous GIC and E-field time-series are used to form a large set of pairwise combinations of \eqref{eqn:gic}. Solving these many linear equations for $\alpha$ and $\beta$ set up ensembles of parameter estimates. The spread in the ensembles is associated directly with accumulated modelling errors. For the ensemble calculation an assumption is required that the network parameters are constant over the time period, as is the case in a resistive network given no change in the network state, e.g. line switching. Such a change would be apparent in modelling as a multi-modal ensemble. Given a time-series $\{t_1,...,t_n\}$, repeated calculation of all ($i,j$) pairs of time instances creates an ensemble of $n(n-1)/2 \approx n^2/2$ (for large $n$) network parameters. Relatively short time periods can by extension result in very large, statistically meaningful ensembles for real-time analysis of the network state, with no actual network information needed. Figure~\ref{fig_ens} is an example of the $\alpha$ ensemble at PAR, with the contributions to the total ensemble from different GIC strengths shown as percentile range profiles.

The ensemble estimation of network parameters presented here has been shown to be stable given different ground conductivity profiles \cite{Heyns2019}. The accuracy of the conductivity profile used significantly influences the spread or error of the ensemble, while factors mentioned above as challenges to traditional GIC modelling also contribute. One result from previous analyses \cite{Wik2008,Heyns2019} is that the empirical network parameters change with GIC intensity. Possible contributions to such observed parameter variation are different parts of a geomagnetic storm having different driving regimes affecting the spatial homogeneity over the network area and possibly the network itself introducing some sort of response. It is possible to model this variation in the form of dynamic network parameters that depend on GIC or E-field magnitude \cite{Heyns2019}. Since the empirical network parameters scale the input E-field to match the measured GIC, the signal-to-noise ratio (SNR) changes as well, adding to the variation of empirical network parameter estimates; in this case the variation is away from a Gaussian noise profile with zero mean associated with GIC at noise levels (as in Figure~\ref{fig_ens}). Although the ensemble method is specific only to a single node-magnetometer pair, the estimated network parameters and spread can be used to calibrate a traditional network model, as is typically used for planning.

\begin{figure}[!ht]
\centering
\includegraphics[width=3.49in]{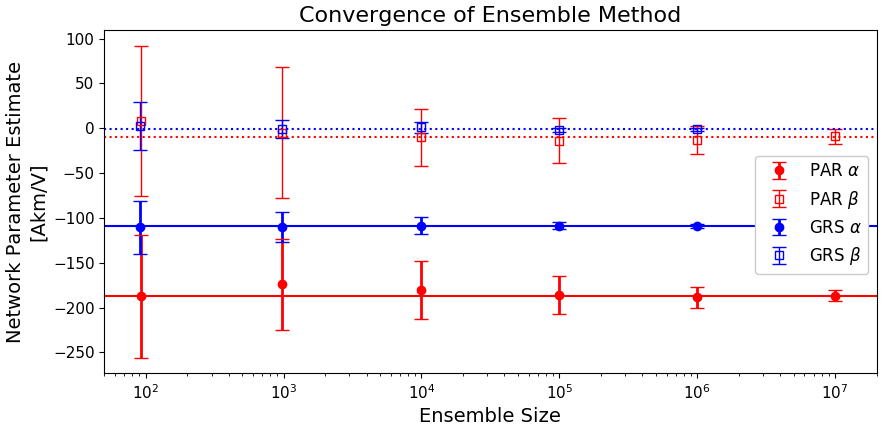}
\caption{Standard deviation and mean of network parameter estimates given different dataset lengths for 2~s (PAR) and 1 minute (GRS) cadence data. Horizontal lines indicate best-fit network parameters, which match the total dataset estimates later in Table \ref{table_para}.}
\label{fig_con}
\end{figure}
\subsection{Convergence and Statistics of Ensemble Method}
Given that limited GIC data has been available in most power networks, there has been a challenge to use data driven modelling in the past. This has changed significantly over the last few years, particularly following the FERC requirement for utilities to measure GICs and make these measurements available \cite{FERC}. 

Nevertheless, the ensemble method used here makes up for the lack of data by creating an ensemble of roughly $n^2/2$ estimates for a dataset of given $n$. Each of these estimates is a possible state of the system, with the peak of the ensemble being the most probable. As the ensembles are heavy-tailed, typical statistics may be misleading. For estimation purposes, the median is inherently robust and the best measure of central tendency. Similarly, the interquartile range (IQR) is an appropriate estimate for the spread in the ensemble. In Figure \ref{fig_con}, the convergence of the ensemble method is shown for 2~s and 1 minute cadence data, given different dataset lengths. For each case, the mean and standard deviation of 25 model runs are plotted, with the best-fit estimate shown as a horizontal line. In each case the final best-fit estimate matches the estimate given by the using the entire training set in a single run. The 1 minute data is smoother and converges more quickly, with variation in estimates being less than 5\% of total network parameters at around $10^5$ estimates. The higher resolution 1~s data, with more variability, has similar adequate convergence at around $10^7$ estimates. This relates to roughly a day's worth of data at either cadence, which means that operationally a single day of data will suffice in characterising the real-time effective state of the network for GIC modelling. Defining the optimal dataset length for convergence furthermore allows the ensemble method to be applied using limited resources - a 2 billion estimate ensemble requires significant computing power for processing. If needed, the optimal dataset lengths can be further reduced by a factor of 10 if multiple runs are averaged such as in Figure \ref{fig_con}.

\section{Data}\label{sec:dat}
In the current work three sets of results, using real world conditions with limited GIC data, no network information, no measured geoelectric field and sparse geomagnetic field measurements (see Table~\ref{table_dat}) are used as an example. Specifically, GIC data from Tennessee, USA (PAR dataset); Queensland, Australia (BOW dataset); and the Eastern Cape, South Africa (GRS dataset) are used. GIC data at all these sites are measured using a Hall-effect sensor at the transformer neutral, with no further data cleaning or filtering besides a 1~s timing drift corrected at PAR. The measured GIC is not the line GIC, but rather the GIC effective to the particular transformer and takes into account the entire network.

Corresponding B-field data provided by the nearest INTERMAGNET (\url{www.intermagnet.org}) station is used to derive either the B-field-to-GIC transfer function or the E-field for the ensemble method. Due to geomagnetic mid-latitude regions being analysed, modelling can be done effectively without spatial interpolation of the B-field to the network in question \cite{Ngwira2009}. Applying an interpolation scheme should decrease the normalised ensemble spread while improving modelling marginally. Such interpolation is sidestepped as the ensemble method directly relates a GIC site with a magnetometer and any error being made is absorbed into the ensemble spreads. For a utility this provides flexibility of not requiring additional magnetometer coverage for operational modelling. The GRS and PAR GIC data have 2~s cadence and BOW has a minimum sampling interval of 4~s during active periods. The associated B-field data for the 2015 datasets are at the modern 1~s cadence, compared to the 2003 dataset which has 60~s B-field data (often still the best cadence many locations have). Appropriate resampling has been done on the concurrent datasets to reflect the lowest cadence for that period.
\begin{table}[!ht]
\renewcommand{\arraystretch}{1.2}
\centering
\caption{Datasets used}
\label{table_dat}
\begin{tabular}{c|ccc}
Dataset & Type & Timespan & Cadence\\
\hline
GRS & GIC & 31/03/2001, 29-31/10/2003 [UTC] & 2 s\\
BOW & GIC & 23/06/2015 [AEST]/[UTC+10] & 4 s\\
PAR & GIC & 22-23/06/2015 [UTC] & 2 s\\
HER & MAG & Same as GRS & 60 s\\
CTA & MAG & Same as BOW & 1 s\\
FRD & MAG & Same as PAR & 1 s
\end{tabular}
\end{table}

For validation, 25\% of each dataset, including the largest GICs associated with the spike-like sudden storm commencement (SSC) and main phase of intense geomagnetic storms, is kept out-of-sample. The models are trained on the remaining lower amplitude GIC data with a lower SNR. Using the `best' data available in each period to train the models on has deliberately been avoided, with the relatively `weak' model results being used to test the model on the validation set. Such an approach with limited data demonstrates conservatively what is possible in real world conditions. Given more data for multiple storms, with more representative training throughout the storm phases, will improve modelling.
\subsection{E-field Derivation}
The ground conductivity (or resistivity) model is an important contribution to the E-field.  Although grounding (earthing) models are widely used in power system analysis, they are usually restricted to power frequency models for fault analysis or high frequency models for lightning studies. GICs are characterized largely by the low frequency spectrum below 50 mHz, for which the skin depth is very deep, conceptually over 50 km \cite{Oyedokun2020}.

Although there are a few sites worldwide where E-field measurements are recorded continuously, it is more typical to derive the E-field from the B-field. Extending the initial assumption that the geomagnetic disturbance field is a plane-wave and that the conductivity of Earth solely depends on depth, we can make use of the basic magnetotelluric (MT) equation \cite{Cagniard1953}. This equation relates the horizontal components of the B-field $B_{x,y}$, to the induced E-field $E_{x,y}$ in the frequency domain,
\begin{align}
\vec{E}(\omega)=\frac{Z(\omega)}{\mu_0}\vec{B}(\omega)\text{ where, }Z(\omega) =   
\left[ \begin{array}{cc} 
Z_{xx}(\omega) & Z_{xy}(\omega) \\ 
Z_{yx}(\omega) & Z_{yy}(\omega)
\end{array} \right].\label{eqn:z}
\end{align}
Here, $Z(\omega)$ is the complex frequency dependent surface impedance, often quoted in units of $[\frac{mV}{km}\frac{1}{nT}]$, i.e. the ratio of electric and magnetic fields in conventional units. The most basic form of the MT equation assumes uniform conductivity. This basic conductivity profile assumption can be extended to a layered-Earth model where conductivity in each layer is assumed to be uniform and defined solely by depth, which produces relative frequency scaling that more accurately defines the Earth's inductive filtering. For both cases, only the off-diagonal impedance tensor components are non-zero, with $Z_{xy}=-Z_{yx}$. When there are more complicated geological structures, such as in coastal regions there is a definite lateral discontinuity or strike, the impedance tensor also becomes more complicated, ultimately with each component unique for a measured surface impedance tensor \cite{Boteler2017}.

Since the ensemble method is stable regardless of the conductivity profile, which is not typically known unless a MT survey has been done, we make use of a representative global average layered-Earth to obtain physically relevant frequency scaling \cite{Sun2015}. This global average profile has layer thicknesses of $d = [40, 210, 160, 260, 230, 1300, 500]$ km and corresponding layer conductivities of $\sigma = [0.0056, 0.0095, 0.0262, 0.0776, 0.526, 1.69, 10]$ S/m. The	conductivity of the terminating half-space is 100 S/m.
\begin{figure}[!ht]
\centering
\includegraphics[width=3.49in]{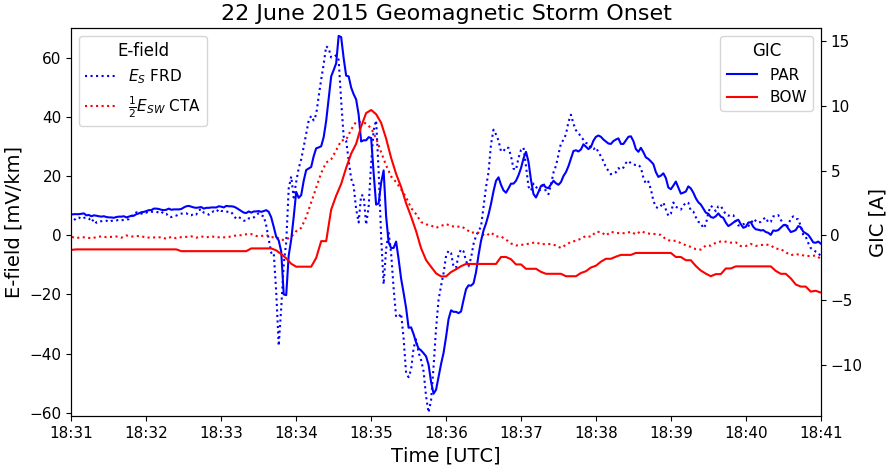}
\caption{Comparison of concurrent GIC measurements in the USA (blue) and Australia (red) with derived E-fields at the sudden storm commencement (SSC). The E-fields are projected onto the typical local network direction, i.e. south for PAR and south-west for BOW. The scaling factor of 1/2 for the CTA E-field matches the E-field axis scaling with measured GIC at BOW. }
\label{fig_SSC}
\end{figure}
\subsection{Data Cadence Implications}
Before modern 1~s cadence B-field and hence E-field data, the resistive approximation was not questioned at all since with 60~s data there is no observable time lag between GIC and E-field. Using the global conductivity profile described above, Figure~\ref{fig_SSC} shows the higher resolution structure of the profiles. An autocorrelation analysis over the geomagnetically active period shows BOW has 12 second delay and PAR has 10 second delay between the E-field and the resultant GIC. By varying the Earth conductivity model, these lags may or may not change, but a contribution from the power network cannot be ruled out. In Figure~\ref{fig_SSC}, which shows the SSC when the initial shock front of solar plasma compresses Earth's B-field and enhances it, followed by an intensification of the magnetospheric ring current which opposes the B-field and weakens it (known as a geomagnetic storm), we also see the lag in B-field and hence E-field between BOW and PAR. For this event PAR was on the Sun-ward side and felt the brunt of the SSC before BOW. At 1~s cadence, the onset of a geomagnetic storm and GIC peak are not concurrent or instantaneous at mid-latitudes globally as is often assumed. As a further illustration of the theoretical basis of \eqref{eqn:trad}, using simple autoscaling of axes and projecting the E-field onto the general network direction to mimic the role of network parameters, the resulting measured GIC and derived general E-field profiles in Figure~\ref{fig_SSC} show strikingly close correlation.
\section{Results}\label{sec:res}
\subsection{Network Parameters and Characterisation}\label{subsec:char}
Besides the time-series modelling capability provided by the ensemble methodology, there is the added impact of finding the effective network response. The resulting effective response requires no prior network modelling, and can absorb the inherent uncertainty in the multi-step modelling. Such uncertainty, seen as deviations from the traditional modelling approach \eqref{eqn:trad}, takes the form of a range (the interquartile range in this case) about the most effective network parameters (summarised in Table~\ref{table_para}). The network parameter range reflects an upper and lower bound for modelling purposes. Given the large number of estimates, the ensembles can be broken down into percentile range profiles associated with GIC strength as seen in Figure~\ref{fig_ens}. These profiles show that the driving regimes for different levels of GICs are often quite different, possibly associated with higher-order effects, and that the empirical estimates are dynamic. Practically, the percentile ranges can be used to further improve modelling, using estimated GIC to define the percentile range regime and update the GIC estimation \cite{Heyns2019}. 
\begin{table}[!ht]
\renewcommand{\arraystretch}{1.2}
\centering
\caption{Site specific network parameters and directionality}
\label{table_para}
\begin{tabular}{c|ccccc}
Site & Estimates & $\alpha$ [A/(V/km)] & $\beta$ [A/(V/km)] & $c$ & Bearing \\
\hline
GRS & $\pm$ 5 million & -109.18 & -1.09 & 0.01  & 181$^\circ$ \\
BOW & $\pm$ 131 million & -211.99 & -167.58 & 0.79 & 218$^\circ$ \\
PAR & $\pm$ 2.1 billion & -185.82 & -12.97 & 0.07 & 184$^\circ$ \\
\end{tabular}
\end{table}

\begin{figure}[!ht]
\centering
\includegraphics[width=3.49in]{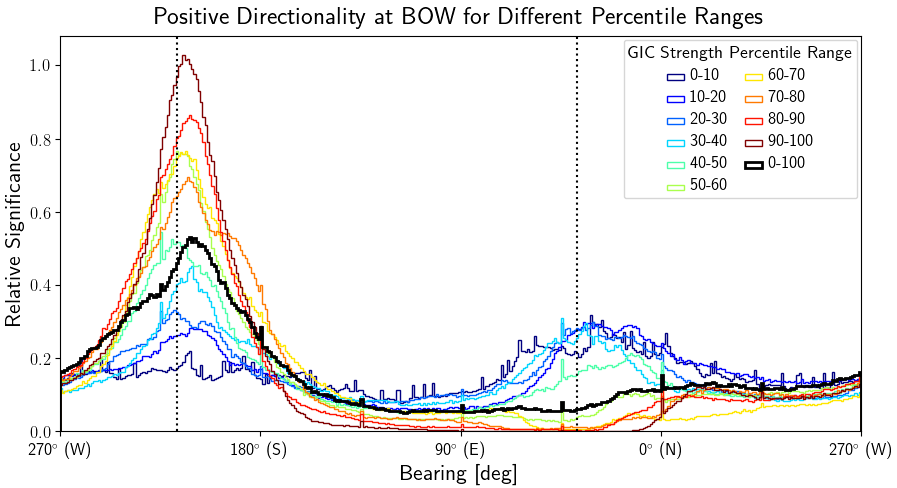}
\caption{The effective ensemble directionality at BOW for different levels of GIC strength along with the bearing of the immediate local line segment indicated by the dashed lines.}
\label{fig_dir}
\end{figure}

From the empirical ensembles, the effective network directionality can also be defined. If there is a strike of sorts in the ground conductivity structure, the effective directionality will absorb it. When rotating a constant B-field, i.e. circular polarisation, a strike results in an elliptical polarised E-field which effectively induces larger GICs along a particular direction. If a simple layered-Earth conductivity model is used, the derived E-field would not have this characteristic and there would be underestimation of modelled E-field along with GIC if the effective network is aligned to the E-field. Using a network parameter ensemble, the additional E-field required for accurate modelling is implicitly encoded in the empirical network parameters and can be seen in the effective network direction. Other non-geophysical contributions to the effective directionality, such as contributions from the network at large, would also be included. The effective directionality argument comes from the arctangent of the ratio of network parameters $c=\beta/\alpha$ \cite{Pulkkinen2007} and is summarised with its associated bearing in Table~\ref{table_para}. The effective directionality can be seen as a planning tool since alignment of the effective weighted network contribution at a node, taking into account the entire network with its shape \cite{Sun2019,Overbye2013}, with the input E-field results in the highest GICs where measured, i.e. the transformers. The relation with actual local transmission line direction between nodes is not as trivial as often made out. An example of a non-trivial directionality ensemble derived using each estimate from the global ensembles is shown for BOW in Figure~\ref{fig_dir}. Although it is the `true' E-field that induces the measured GICs, with the ensemble method, given a consistent system state, we have a direct association between an approximate input E-field direction and high transformer GICs.
\subsection{Empirical Transfer Functions}\label{subsec:tf}
\begin{figure*}[!t]
\centering
\hfil%
{\includegraphics[width=7.0in]{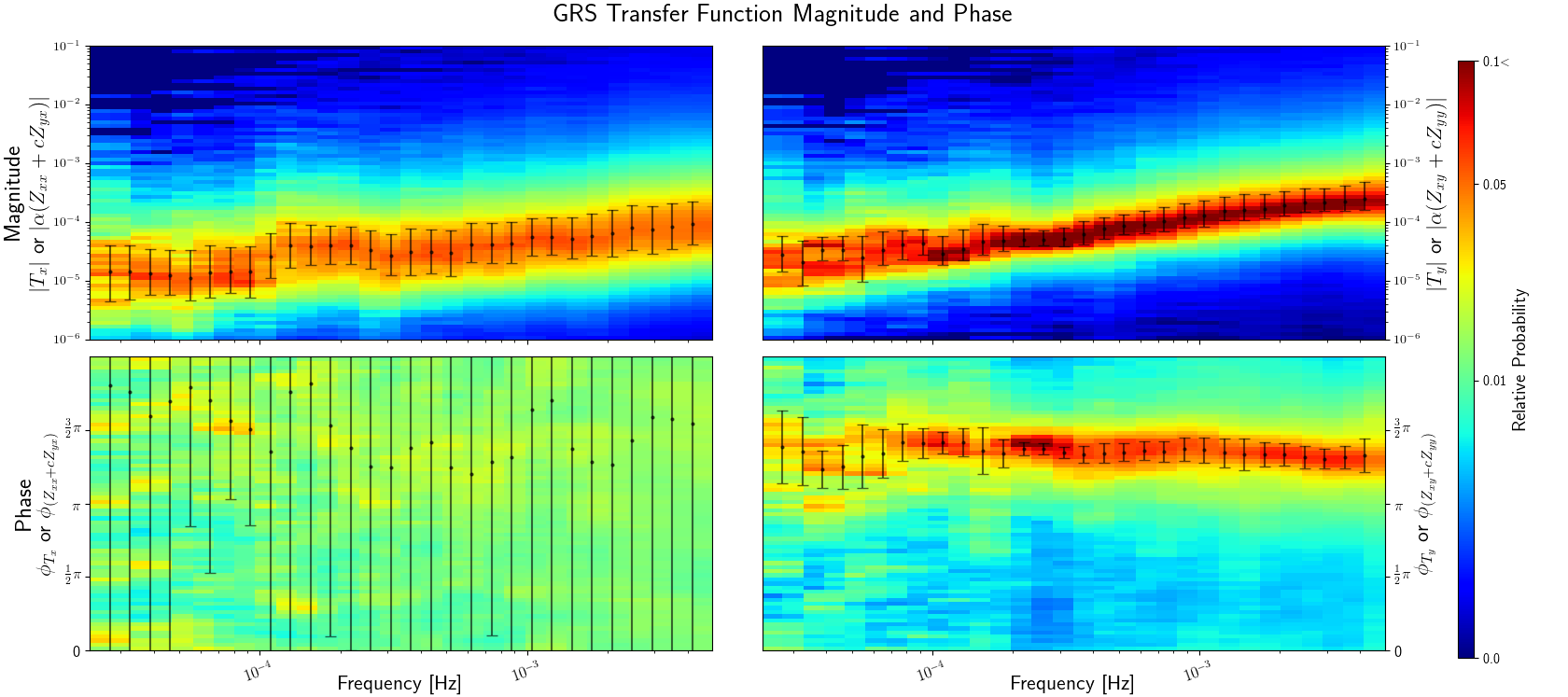}}%
\hfil
\caption{Transfer function response at GRS, with magnitude and phase of the $T_{x,y}$ components shown. Since the effective network orientation is north-south and $c\approx0$, only the single off-diagonal magnetotelluric impedance $Z_{xy}$ dominates the responses, with $Z_{yx}$ (and hence $T_x$) being swamped by noise.}
\label{fig_qimp}
\end{figure*}
It is also possible to directly relate the B-field to GIC via an empirical transfer function. Through this process the ground conductivity assumption required to derive the E-field, which introduces significant error, is bypassed. The derived transfer function (TF) is the net effective transform of the Earth over the entire network area and any power network response. The B-field-to-GIC TF approach has been followed before more from a traditional MT background, without decomposing the network filtering \cite{Ingham2018}. Although the results have shown good correspondence with known geophysical features, there are cases where such interpretations failed. When these interpretations fail, it can largely be attributed to the results of a GIC based TF taking into account the network. To take into account the effect of the network we firstly, update \eqref{eqn:gic} with $c=\beta/\alpha$. The matrix form of this equation becomes,
\begin{align}
\Gamma(t)& = \alpha E_x(t) + \alpha cE_y(t) \nonumber
\\& = \begin{matrix}\begin{bmatrix} \alpha & \alpha c \end{bmatrix}\\\mbox{}\end{matrix} 
\left[ \begin{array}{c} E_x(t) \\ E_y(t) \end{array} \right].
\end{align}
In the frequency domain, we use measured B-field data without the need for an assumed conductivity related impedance tensor and the uncertainty that goes with it,
\begin{align}
\Gamma(\omega)=& \begin{matrix}\begin{bmatrix} \alpha & \alpha c \end{bmatrix}\\\mbox{}\end{matrix} 
\left[ \begin{array}{c} E_x(\omega)\\ E_y(\omega)\end{array} \right] \nonumber 
\\=& \begin{matrix}\begin{bmatrix} \alpha & \alpha c \end{bmatrix}\\\mbox{}\end{matrix} 
\begin{bmatrix} Z_{xx}(\omega) & Z_{xy}(\omega) \\ Z_{yx}(\omega) & Z_{yy}(\omega) \end{bmatrix} 
\left[ \begin{array}{c} B_x(\omega) \\ B_y(\omega) \end{array} \right] \nonumber
\\=& \left[ \begin{array}{c} \alpha\big(Z_{xx}(\omega)+cZ_{yx}(\omega)\big) \\ \alpha\big(Z_{xy}(\omega)+cZ_{yy}(\omega)\big) \end{array} \right]^T\left[ \begin{array}{c} B_x(\omega) \\ B_y(\omega) \end{array} \right] \nonumber
\\=& \begin{matrix}\begin{bmatrix} T_x(\omega) & T_y(\omega) \end{bmatrix}\\\mbox{}\end{matrix}  \left[ \begin{array}{c} B_x(\omega) \\ B_y(\omega) \end{array} \right] \label{eqn:tf1}
\\=&\mathbf{TF}_{B_x}(\omega)+\mathbf{TF}_{B_y}(\omega).\label{eqn:tf}
\end{align}
In the formalism above, $T_{x,y}(\omega)$ in \eqref{eqn:tf1} are the components of the B-field-to-GIC TF. $\mathbf{TF}_{B_x,B_y}(\omega)$ in \eqref{eqn:tf} are not TF components, but rather the result of multiplying $T_{x,y}(\omega)$ with their associated B-field components. To aid in identifying effective contributions of the surface impedance tensor \eqref{eqn:z} in the TF, the geometric scaling $c$ is split from the network parameters. The final TF absorbs all network filtering of $\alpha$ (which cannot be separated) and can be directionally limited given certain effective network topologies (when $c$ or $c^{-1}\approx0$), severely affecting the SNR. This is seen in  Figure~\ref{fig_qimp} for the GRS TF; the north-south effective orientation of the network dampens $T_x$, and by extension the contribution from $B_x$. In a resistive network, $\alpha$ is assumed to be constant. This may not be the case, and $\alpha$ may in fact have frequency dependent scaling \cite{Weigel2019}.

The TF approach has a number of inherent advantages:
\begin{itemize}
\setlength\itemsep{0.2em}
\item[a.] The measured B-field is directly related to measured GIC without any further assumptions of ground or network response. Historically, the B-field is much more commonly measured and better understood than the E-field. The long term B-field analyses allow for better extreme value statistical modelling important in planning.
\item[b.] The integrated ground and network effects of the entire network are modelled for the node. Any variation in these responses are encoded in the TF spread.
\item[c.] Different regimes of driving geomagnetic disturbances have different spectral components, which are accurately modelled in the frequency domain along with geophysical anomalies and network responses. The ensemble method requires further processing for similar modelling, and failing that can only increase its uncertainty spread.
\item[d.] The TF estimates are statistically and physically meaningful, i.e. they allow for probabilistic GIC estimation and interpretation of the geophysical and network contributions to the TF.
\end{itemize}
Shortcomings of using a TF approach are the need for representative training data and correlation in time between geomagnetic disturbances at the B-field measurement site and over the effective network. Both these issues are apparent at PAR (discussed in \ref{sec:mod}), in which case the more robust E-field ensemble approach does better.

In order to estimate the TF, an ensemble method in the frequency domain is used. For a number of windows, the two complex TF components are estimated at each frequency through many simultaneous equations calculations (negative frequencies are folded). These ensembles are then log-binned and the median and interquartile range estimated for modelling purposes as with the previous ensemble method. The limited data in this work is a challenge since there may well be different spectra to consider and the SNR is poor at many frequencies. As a result only the GRS and PAR datasets were used for TF modelling. The GRS TF used an 18 hour window, with 3 hour shifting, at 60~s sampling and the PAR TF used a 6 hour window, with 2 hour shifting at 2~s sampling. Both dataset processing schemes give a few thousand estimates per frequency, which is then significantly increased with log-binning. However, at PAR the short window length may have missed certain typical lower frequencies such as daily variation. Given a longer dataset it can be assumed that the TF will become significantly more accurate. 

\begin{figure}[!ht]
\centering
\includegraphics[width=3.49in]{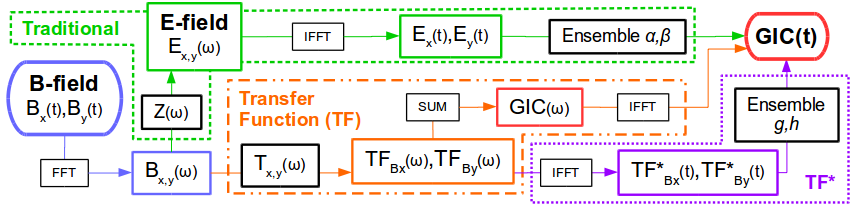}
\caption{Flow of the traditional ensemble updated layered-Earth (LE), transfer function (TF) and ensemble updated transfer function (TF*) models used.}
\label{fig_mod}
\end{figure}

Uncertainty in the time domain reconstruction is not trivial since the frequency specific uncertainty bounds can occur in any combination, constructively or destructively. To obtain an uncertainty estimate in the time domain, the time domain ensemble method from earlier is reintroduced to generate time domain scaling and uncertainty bands. Taking the IFFT of the TF result $\mathbf{TF}_{B_x,B_y}(\omega)$ in \eqref{eqn:tf}, we can use time domain version of the same result (denoted by TF*) analogously to the E-field-to-GIC ensemble methodology,
\begin{align}
\Gamma(t)=g\mathbf{TF}^*_{B_x}(t)+h\mathbf{TF}^*_{B_y}(t). \label{eqn:gh}
\end{align}
The $g$ and $h$ parameters from \eqref{eqn:gh} generate an uncertainty spread and improve modelling by tuning TF* or the analogous `network effective E-field' in the time domain. Here the $g$ and $h$ ensembles are centred roughly around 1, which is expected since the empirical TF should correctly scale the input B-field to GIC. Any significant deviation from unity scaling suggests the TF does not adequately link B-field to GIC, such as when the separation between magnetometer and substation is too great. For reference, the separation of PAR-FRD pair is extreme at around 1000 km whereas the GRS-HER pair is roughly 700 km apart. When further scaling is needed, the ensemble TF can further improve modelling. In either case of an ideal TF with unity scaling or tweaked TF with further scaling, the spread of the $g$ and $h$ parameters allow for accurate uncertainty estimation in the time domain and not the frequency domain where the TF is defined. The conceptual flow of the B-field-to-GIC TF and TF* ensembles and E-field ensembles are summarised in Figure~\ref{fig_mod}.
\subsection{GIC Modelling}\label{sec:mod}
The results of the layered-Earth (LE) ensemble, TF and TF* models discussed are summarised in Table~\ref{table_mod}. Of the measures used, the RMSE and correlation coefficient $\rho$ are typical. The relative error (RE) has been used in previous studies and quantifies the percentage error made in terms of the signal amplitude. Considering the GIC sites, only GRS \cite{Matandirotya2015} and BOW \cite{Marshall2019} have previously been modelled at all and only GRS has comparable performance metrics for the validation set. In previous modelling the best result at GRS for the same validation set used finite element modelling, obtaining a RMSE of 0.98A and RE of 42\% compared to 0.69A and 25.1\% using the TF* model in this work. 
\begin{table}[!ht]
\renewcommand{\arraystretch}{1.2}
\centering
\caption{Modelling results for out-of-sample data}
\label{table_mod}
\begin{tabular}{c|ccc}
Site (Model) & RMSE (A) & $\rho$ & RE (\%) \\
\hline
GRS (LE) & 0.91 & 0.86 & 32.3 \\
BOW (LE) & 1.24 & 0.92 & 32.0 \\
PAR (LE) & 1.23 & 0.89 & 44.6 \\
GRS (TF) & 0.81 & 0.86 & 29.4 \\
GRS (TF*) & 0.69 & 0.91 & 25.1 \\
PAR (TF) & 1.49 & 0.81 & 57.3 \\
PAR (TF*) & 1.32 & 0.86 & 52.5
\end{tabular}
\end{table}

When dealing with higher amplitude GIC and associated higher amplitude noise, such as at PAR, the RE measure can be significantly skewed. BOW and GRS have similar magnitude GICs, making their RE results more comparable. The RE measure can be skewed even further with limited data used for modelling; the dynamic parameters used may change significantly given a geomagnetic storm profile or amplitude different (usually higher) to data previous trained on. This is seen in the current work as an underestimation in modelling particularly for BOW, where the out-of-sample SSC is significantly higher than anything in the training set. 

\begin{figure}[!ht]
\centering
\includegraphics[width=3.49in]{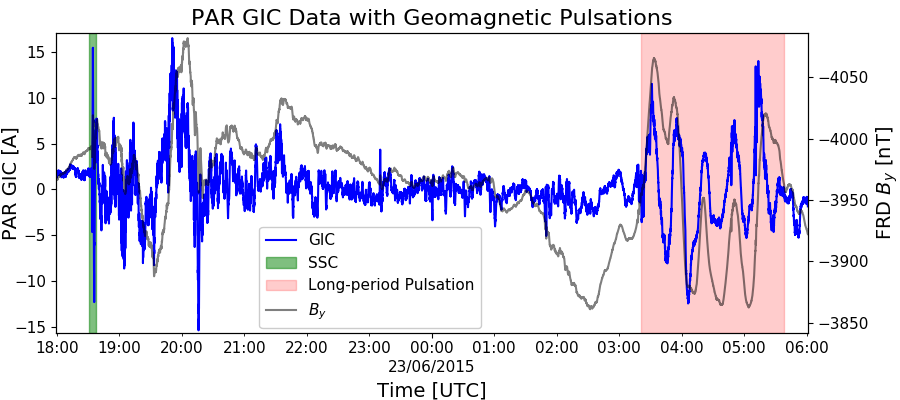}
\caption{The PAR GIC time-series, with the dominant FRD $B_y$ component ($c\approx0$), shows significant driving from long-period geomagnetic pulsations resulting in sustained GICs of similar amplitude to the SSC spike.}
\label{fig_gicpara}
\end{figure}

\begin{figure*}[!t]
\centering
\hfil%
{\includegraphics[width=7.0in]{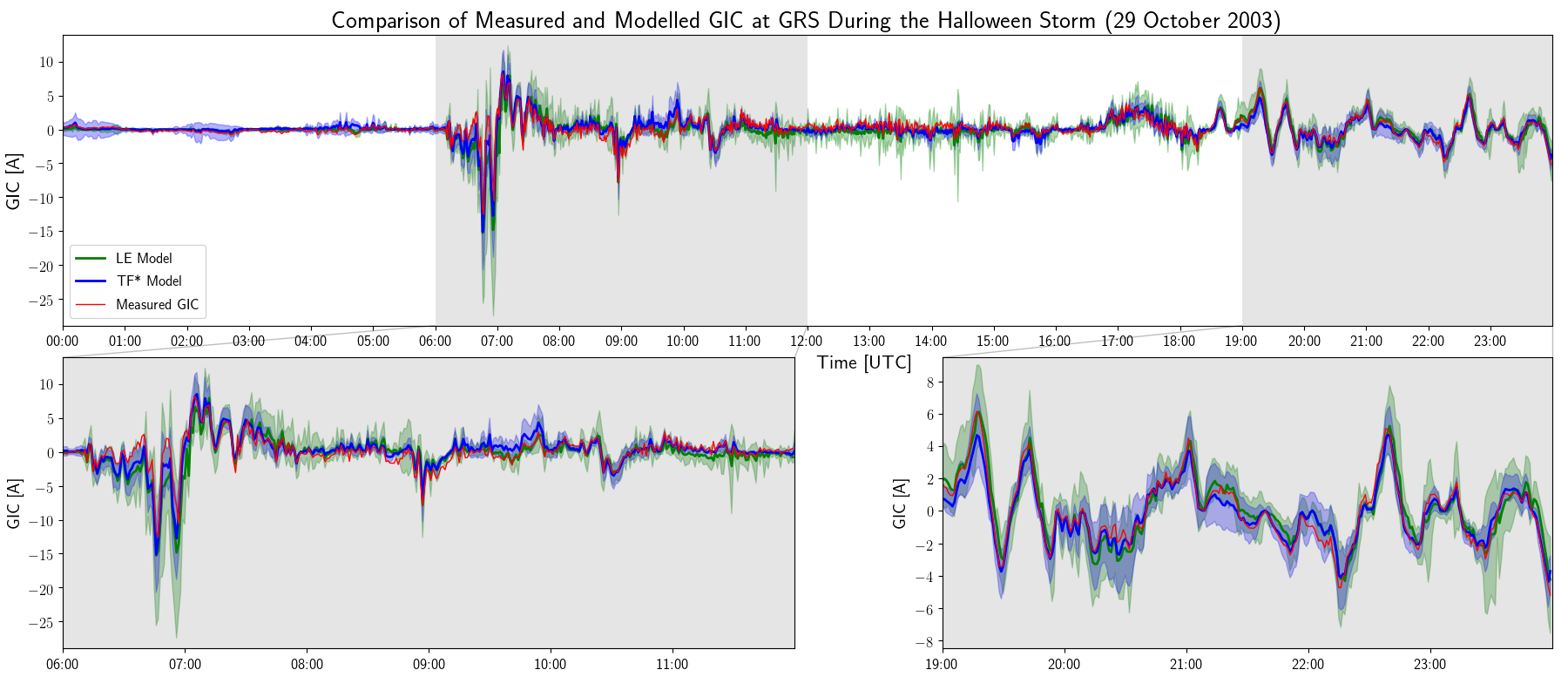}}%
\hfill
\caption{Time-series of the resulting GIC modelling at GRS for the out-of-sample dataset, including the associated uncertainty bands. More than 80\% of data points fall within the TF* prediction band compared to 70\% for the LE prediction band.}
\label{fig_band}
\end{figure*}

PAR presents a further modelling challenge since the geomagnetic field used in training has a relatively high content of long-period (20 min) geomagnetic pulsations that are usually associated with geomagnetic substorms and auroral regions \cite{Saito1978}. During a 2 hour period, the pulsations caused high amplitude GICs, comparable to the SSC peak GIC (see Figure~\ref{fig_gicpara}). These driving pulsations are interesting in two respects:
\begin{itemize}
\setlength\itemsep{0.2em}
\item[a.] They indicate geomagnetic activity associated more often with auroral regions. Here significant effects are seen in what is generally considered to be a mid-latitude region, suggesting the spatial distribution models for geomagnetic disturbances need to be reviewed.
\item[b.] They are non-stationary and localised, which violates the plane-wave assumption of the disturbance B-field. Due to the separation between sites, during the period of pulsations there is an offset between the measured B-field at FRD and the B-field driving the GIC at PAR. Although the offset distorts the models, particularly the TF's, inspection shows the results are useful. In practice, with more data being accumulated and spatial interpolation, the distortion seen in modelling may be minimised.
\end{itemize}
Nevertheless, the presented modelling is viable from an operational outlook especially with the uncertainty band tracking the error introduced, as shown for GRS in Figure~\ref{fig_band}.
\section{Conclusion and Further Research}
The ensemble methodology has shown how the relatively simple governing nodal GIC equation can be leveraged given measured data to represent much more complicated GIC modelling dynamics - even with relatively few, spatially sparse, measurement sites over a large area and limited temporal coverage. The usefulness of the methodology is increased by uncertainty being included, even in cases of extreme separation between B-field and GIC measurement. In a utility, the ensemble and TF methods can be used in control centres during or before a geomagnetic disturbance, giving direct indication of simulated GIC exposure of several transformers (providing there have been prior measurements of GIC) and identifying discrepancies in state of the network, all without overloading the communication systems. In this regard the data driven methods already demonstrate the effectiveness of FERC order 830 for the collection of data. Under this directive, it is foreseeable that data driven methods become even more viable. The associated B-field measurements considered do not need 1~s cadence to be operationally effective, although they do allow for high resolution GIC dynamics to be seen, such as SSC propagation across the globe. From a planning perspective, the ensemble method can be used to test transformer level models of uneven GIC distributions between transformers in the same substation \cite{Divett2018} and calibrate general network models further. The significant impact of driving geomagnetic pulsations does however suggest that the definition of mid-latitude GIC drivers and the plane-wave spatial scale be reconsidered; given the expansion of high-latitude effects, the currently understood risk to mid-latitude power systems may be underestimated. 
%
%

\begin{thebibliography}{23}

\bibitem{Albertson1970}
V.~Albertson and J.~{Van Baelen}, ``{Electric and Magnetic Fields at the Earth's Surface Due to Auroral Currents},'' \emph{IEEE Transactions on Power Apparatus and Systems}, vol. PAS-89, no.~4, pp. 578--584, 1970.

\bibitem{Lehtinen1985a}
M.~Lehtinen and R.~Pirjola, ``{Currents produced in earthed conductor networks by geomagnetically-induced electric fields},'' \emph{Annales Geophysicae}, vol.~3, no.~4, pp. 479--484, 1985.

\bibitem{Sun2019}
R.~Sun and C.~Balch, ``{Comparison between 1-D and 3-D Geoelectric Field Methods to Calculate Geomagnetically Induced Currents: A Case Study},'' \emph{IEEE Transactions on Power Delivery}, vol.~34, no.~6, pp. 2163--2172, 2019.

\bibitem{Weigel2019}
R.~S. Weigel and P.~J. Cilliers, ``{An Evaluation of the Frequency Independence Assumption of Power System Coefficients Used in Geomagnetically Induced Current Estimates},'' \emph{Space Weather}, vol.~17, no.~12, pp. 1674--1688, 2019.
  
\bibitem{Lucas2018}
G.~M. Lucas, J.~J. Love and A.~Kelbert, ``{Calculation of Voltages in Electric Power Transmission Lines During Historic Geomagnetic Storms: An Investigation Using Realistic Earth Impedances},'' \emph{Space Weather}, vol.~16, no.~2, pp. 185--195, 2018.
  
\bibitem{Overbye2013}
T.~J. Overbye, K.~S. Shetye, T.~R. Hutchins, Q.~Qiu, and J.~D. Weber, ``{Power Grid Sensitivity Analysis of Geomagnetically Induced Currents},'' \emph{IEEE Transactions on Power Systems}, vol.~28, no.~4, pp. 4821--4828, 2013.

\bibitem{Divett2018}
T.~Divett et al., ``{Transformer-Level Modeling of Geomagnetically Induced Currents in New Zealand's South Island},'' \emph{Space Weather}, vol.~16, no. 6, pp. 1--18, 2018.

\bibitem{Chisepo}
H.~K. Chisepo, C.~T. Gaunt and L. D. Borrill, ``{Measurement and FEM analysis of DC/GIC effects on transformer magnetization parameters},'' \emph{2019 IEEE Milan PowerTech}, pp. 1--6, 2019. doi:10.1109/PTC.2019.8810423
  
\bibitem{Chisepo2}
P. Jankee et al., ``{Transformer models and meters in MATLAB and PSCAD for GIC and leakage dc studies},'' \emph{2020 IEEE International SAUPEC/RobMech/PRASA Conference}, pp. 1--6, 2020. doi:10.1109/SAUPEC/RobMech/PRASA48453.2020.9041060

\bibitem{Blake2018b}
S.~P. Blake et al., ``{A Detailed Model of the Irish High Voltage Power Network for Simulating GICs},'' \emph{Space Weather}, vol.~16, no.~11, pp. 1770--1783, 2018.

\bibitem{Wik2008}
M.~Wik et al., ``{Calculation of geomagnetically induced currents in the 400 kV power grid in southern Sweden},'' \emph{Space Weather}, vol.~6, no.~7, pp. 1--11, 2008.

\bibitem{Heyns2019}
M.~J. Heyns, S.~I. Lotz, P.~J. Cilliers and C.~T. Gaunt, ``Ensemble Estimation of Network Parameters: A Tool to Improve the Real-time Estimation of GICs in the South African Power Network,'' in \emph{The Proceedings of SAIP2017, the 62nd Annual Conference of the South African Institute of Physics}, edited by J. Engelbrecht, Stellenbosch, pp. 270--275, 2017. Available online at http://events.saip.org.za
 
\bibitem{FERC}
``Federal Energy Regulatory Commission: Reliability Standard for Transmission System Planned Performance for  Geomagnetic Disturbance Events.'' Order 830, Sep 2016. Washington DC.

\bibitem{Ngwira2009}
C.~M. Ngwira et al., ``{Limitations of the modeling of geomagnetically induced currents in the South African power network},'' \emph{Space Weather}, vol.~7, no.~10, pp. 1--5, 2009.

\bibitem{Oyedokun2020}
D. Oyedokun, M. J. Heyns, P. J. Cilliers and C. T. Gaunt, ``{Frequency Components of Geomagnetically Induced Currents for Power System Modelling},'' \emph{2020 IEEE International SAUPEC/RobMech/PRASA Conference}, pp. 1--6, 2020. doi:10.1109/SAUPEC/RobMech/PRASA48453.2020.9041021

\bibitem{Cagniard1953}
L.~Cagniard, ``{Basic Theory Of The Magneto‐Telluric Method Of Geophysical Prospecting},'' \emph{Geophysics}, vol.~18, no.~3, pp. 605--635, 1953.

\bibitem{Boteler2017}
D.~H. Boteler and R.~J. Pirjola, ``{Modeling geomagnetically induced currents},'' \emph{Space Weather}, vol.~15, no.~1, pp. 258--276, 2017.

\bibitem{Sun2015}
J.~Sun, A.~Kelbert and G.~D. Egbert, ``{Ionospheric current source modeling and global geomagnetic induction using ground geomagnetic observatory data},'' \emph{Journal of Geophysical Research: Solid Earth}, vol.~120, no.~10, pp. 6771--6796, 2015.

\bibitem{Pulkkinen2007}
A.~A. Pulkkinen, R.~Pirjola, and A.~Viljanen, ``{Determination of ground conductivity and system parameters for optimal modeling of geomagnetically induced current flow in technological systems},'' \emph{Earth, Planets and Space}, vol.~59, no.~9, pp. 999--1006, 2007.

\bibitem{Ingham2018}
M.~Ingham et al., ``{Assessment of GIC based on transfer function analysis},'' \emph{Space Weather}, vol.~15, no.~12, pp. 1615--1627, 2017.

\bibitem{Matandirotya2015}
E.~Matandirotya, P.~J. Cilliers and R.~R. Van Zyl, ``{Modeling geomagnetically induced currents in the South African power transmission network using the finite element method},'' \emph{Space Weather}, vol.~13, no.~3, pp. 185--195, 2015.

\bibitem{Marshall2019}
R.~A.~Marshall et al., ``{Modelling Geomagnetically Induced Currents in Australian power networks using different conductivity models},'' \emph{Space Weather}, vol.~17, no.~5, pp. 727-756, 2019.

\bibitem{Saito1978}
M. J. Heyns, S. I. Lotz and C. T. Gaunt, ``{Geomagnetic Pulsations Driving Geomagnetically
Induced Currents},'' \emph{ESSOAr} (preprint on https://essoar.org/), 2020. doi:10.1002/essoar.10503394.1

\end{thebibliography}
%
%
%

%
\end{document}